\documentclass[aps,prc,preprint,showpacs,superscriptaddress,amsmath,amssymb,floatfix,nofootinbib]{revtex4}

\usepackage[pdftex]{graphicx}
\usepackage{dcolumn}
\usepackage{bm}

\def\pmb#1{\setbox0=\hbox{#1}%
  \kern-.025em\copy0\kern-\wd0 
  \kern.05em\copy0\kern-\wd0
  \kern-.025em\raise.0433em\box0 }

\def\lambdabar{\protect\@lambdabar}
\def\@lambdabar{%
\relax
\bgroup
\def\@tempa{\hbox{\raise.73\ht0
\hbox to0pt{\kern.25\wd0\vrule width.5\wd0
height.1pt depth.1pt\hss}\box0}}%
\mathchoice{\setbox0\hbox{$\displaystyle\lambda$}\@tempa}%
{\setbox0\hbox{$\textstyle\lambda$}\@tempa}%
{\setbox0\hbox{$\scriptstyle\lambda$}\@tempa}%
{\setbox0\hbox{$\scriptscriptstyle\lambda$}\@tempa}%
\egroup
}

\begin{document}

\preprint{J-PARC-TH-0333}

\title{\boldmath
Sensitivity of the $^{3,4}$He($K^-$, $\pi^0$) production ratio to 
the $\Lambda$ binding energy of $^3_\Lambda$H 
}

\author{Toru~Harada}%
\email{harada@osakac.ac.jp}
\affiliation{%
Center for Physics and Mathematics,
Osaka Electro-Communication University, Neyagawa, Osaka, 572-8530, Japan
}
\affiliation{%
J-PARC Branch, KEK Theory Center, Institute of Particle and Nuclear Studies,
High Energy Accelerator Research Organization (KEK),
203-1, Shirakata, Tokai, Ibaraki, 319-1106, Japan
}

\author{Yoshiharu~Hirabayashi}%
\affiliation{%
Information Initiative Center, 
Hokkaido University, Sapporo, 060-0811, Japan
}

\date{\today}

\begin{abstract}
We study the production of $^3_\Lambda$H and $^4_\Lambda$H
in the $^{3,4}$He($K^-$,$\pi^0$) reactions at
$p_{K^-}=1.0$~GeV/$c$ within the distorted-wave impulse approximation,
using the optimal Fermi-averaged $K^-p\to\pi^0\Lambda$ amplitude.
Because the $^3_\Lambda$H ground state is extremely weakly bound,
the $d$--$\Lambda$ wave function becomes spatially extended.
We calculate the integrated cross sections $\sigma_{\rm lab}$ and their ratio
$R_{34}=\sigma_{\rm lab}(^3_\Lambda{\rm H})/\sigma_{\rm lab}(^4_\Lambda{\rm H})$
for forward angles $\theta_{\rm lab}=0^\circ$--$20^\circ$.
The production strength of $^3_\Lambda$H and the ratio $R_{34}$ are
strongly sensitive to the $\Lambda$ binding energy $B_\Lambda$, 
which is constrained to be approximately 0.05--0.15~MeV
by comparison with experimental data from the J-PARC E73 experiment.
This indicates that the $^3$He($K^-$,$\pi^0$) reaction provides
a sensitive probe of the weak binding of $^3_\Lambda$H.
\end{abstract}
\maketitle


\section{\boldmath
Introduction}

The hypertriton $^3_\Lambda$H is the lightest $\Lambda$ hypernucleus and 
also one of the most delicate bound states in nuclear physics~\cite{Juric73}.
Its $\Lambda$ binding energy $B_\Lambda = M(d)+m_\Lambda - M(^3_\Lambda{\rm H})$ 
is extraordinarily small on the nuclear scale,
so that the $\Lambda$ wave function inevitably develops a long tail.
This feature makes $^3_\Lambda$H a stringent benchmark for the $\Lambda N$ interaction
near threshold and amplifies the impact of $B_\Lambda$ on observables that probe
the spatial overlap with a compact nuclear core.

Recently, the $^3_\Lambda$H ground state (g.s.) has drawn renewed attention,
partly because its weak binding is intertwined with the mesonic weak-decay
and lifetime issue~\cite{HypHI13,Adam20,Acharya23,Kasagi25},
and partly because \textit{ab initio} calculations based on modern $NN$ and $YN$ interactions
have reached a level where sensitivity to long-range dynamics can be quantified
in a controlled manner~\cite{Gal19,Haidenbauer23,Kamada24}.

However, the precise value of the $\Lambda$ binding energy
$B_\Lambda$ in $^3_\Lambda$H has not yet been established conclusively.
Several determinations of $B_\Lambda(^3_\Lambda{\rm H})$ show a certain spread,
including weak-binding values from reaction analyses~\cite{Kasagi25,Akaishi26},
as well as larger values reported by the STAR Collaboration~\cite{Adam20}
and by decay-pion spectroscopy at MAMI~\cite{Kino25}.

Production reactions provide such an opportunity to probe this feature.
In particular, the $^3$He($K^-$,$\pi^0$) reaction at forward angles
is driven predominantly by the non-spin-flip $\Delta S=0$ elementary process
$K^-p\to\pi^0\Lambda$.
Under these kinematics the reaction directly probes the spatial overlap
between the initial proton distribution in $^3$He and the final $\Lambda$
distribution in $^3_\Lambda$H.
Consequently, if $^3_\Lambda$H is as loosely bound as suggested 
by emulsion~\cite{Juric73,Kasagi25}
and modern analyses~\cite{HypHI13,Adam20,Acharya23,Akaishi26,Kino25},
its extended wave function should leave a visible imprint on the absolute yield and,
more clearly, on suitable cross-section ratios.

In this paper, 
we investigate the sensitivity of the $^3_\Lambda$H
production strength to the $\Lambda$ binding energy $B_\Lambda$
within the distorted-wave impulse approximation (DWIA),
employing the optimal Fermi-averaged $K^-p\to\pi^0\Lambda$ amplitude.
To reduce common systematic uncertainties,
we focus on the ratio of integrated cross sections
\begin{equation}
R_{34}=\sigma_{\rm lab}(^3_\Lambda{\rm H})/\sigma_{\rm lab}(^4_\Lambda{\rm H}),
\label{eqn:e0}
\end{equation}
where $^4_\Lambda$H is produced under closely related conditions
within the same theoretical framework.
We calculate the integrated cross sections $\sigma_{\rm lab}$ 
and the ratio $R_{34}$ for forward angles,
and examine how the experimental value $R^{\rm exp}_{34}$
reflects the spatial extension of the $\Lambda$ wave function
in $^3_\Lambda$H.
Unlike our previous studies employing empirical binding energies~\cite{Harada21,Harada19},
the present work explicitly quantifies the $B_\Lambda$ dependence
of both $\sigma_{\rm lab}(^3_\Lambda{\rm H})$ and $R_{34}$.

\section{\boldmath
Methodology}

We describe the $^{3,4}$He($K^-$,$\pi^0$)$^{3,4}_\Lambda$H reactions 
at $p_{K^-}=$ 1.0 GeV/$c$ in the DWIA framework~\cite{Harada21,Harada19}
with the optimal Fermi-averaging~\cite{Harada04}, using 
the $K^-p\to\pi^0\Lambda$ elementary amplitudes given by Gopal et al.~\cite{Gopal77}.
At forward angles the dominant contribution comes from the non-spin-flip
$\Delta S=0$ elementary process $K^-p\to\pi^0\Lambda$ embedded in the nucleus.
Following the effective-number approach~\cite{Hufner74,Dover80},
the laboratory differential cross section for producing 
a bound hypernuclear state $J_B^P$ is written as
\begin{eqnarray}
\left({d\sigma \over d\Omega}\right)_{\rm lab, \theta_{\rm lab}}^{J_B^P}
= \alpha \left\langle{\frac{d\sigma}{d\Omega}}
\right\rangle^{\rm elem}_{\rm lab, \theta_{\rm lab}}
Z^{J_B^P}_{\rm eff}(\theta_{\rm lab}),
\label{eqn:e1}
\end{eqnarray}
where 
$\alpha\langle d\sigma/d\Omega\rangle^{\rm elem}_{\rm lab}=
\alpha|\overline{f}_{K^-p\to\pi^0\Lambda}|^2$
represents the in-medium ${K^-}p\to\pi^0\Lambda$ cross section 
including the kinematical factor $\alpha$~\cite{Dover83}.
The effective number of protons~\cite{Harada19} is expressed as
\begin{eqnarray}
Z^{J_B^P}_{\rm eff}(\theta_{\rm lab}) = C^2_{TS}|F(q)|^2,
\label{eqn:e2}
\end{eqnarray}
where $C_{TS}$ is the isospin-spin spectroscopic amplitude.
The form factor is given by
\begin{eqnarray}
F(q)=\int_0^{\infty}dr\, r^2\,\rho_{\rm tr}(r)\,
\widetilde{j}_{0}\!\left(q;\frac{M_C}{M_D}r\right),
\label{eqn:e3}
\end{eqnarray}
with the distorted-wave function $\widetilde{j}_{0}(q;r)$ describing meson distortion, 
where $M_D\equiv(M_B+M_A)/2$; $M_A$, $M_B$, and $M_C$ are masses of the 
target nucleus, the hypernucleus, and the nuclear core, respectively.
For very light systems, the recoil effect is not a small correction but a kinematical necessity.
In the eikonal approximation, we incorporate recoil by scaling the coordinate as
$r\to (M_C/M_D)r$. 
This leads naturally to the effective momentum transfer
$q_{\rm eff}\equiv(M_C/M_D)q\simeq[(A-1)/A]q$,
which controls how recoil reshapes the production strength~\cite{Dover80,Auerbach83}.
In the plane-wave limit (distortion parameters $\sigma_{K^-},\sigma_\pi\to0$), 
one recovers PWIA by replacing
$\widetilde{j}_0(q;r)$ with the spherical Bessel function $j_0(qr)$~\cite{Harada19}.

The transition density entering Eq.~(\ref{eqn:e3}) is
\begin{eqnarray}
\rho_{\rm tr}(r)=\varphi^{(\Lambda)*}_{0}(r)\,\varphi^{(N)}_{0}(r),
\label{eqn:e4}
\end{eqnarray}
where $\varphi^{(\Lambda)}_{0}=\langle\phi^{(C)}_0|\Psi_B\rangle$ is the relative wave function
(spectroscopic amplitude) for $\Lambda$ in ${^{3,4}_\Lambda{\rm H}}$~\cite{Kurihara85},
and $\varphi^{(N)}_{0}=\langle\phi^{(C)}_0|\Psi_A\rangle$ is the corresponding nucleon 
wave function in ${^{3,4}{\rm He}}$~\cite{Akaishi86}.
For $A=4$, we employ the $3N$--$\Lambda$ model based on four-body 
$\Lambda NNN$ calculations~\cite{Harada19},
reproducing $B_\Lambda=2.16$~MeV~\cite{Schulz16} and 
the root-mean-square (r.m.s.) $t$--$\Lambda$ distance 
$\langle r^2_{t\mbox{-}\Lambda} \rangle^{1/2}=3.68$~fm 
for ${^4_\Lambda{\rm H}}$($J^P=0^+$, g.s.),
together with $\varphi^{(N)}_0$ from four-body $NNNN$ calculations~\cite{Akaishi86} 
and $C^2_{TS}=2$ in Eq.~(\ref{eqn:e2}).
For $A=3$, we employ the $2N$--$\Lambda$ model~\cite{Harada15a} based 
on microscopic CDCC calculations~\cite{Kamimura86},
reproducing $B_\Lambda=0.13$~MeV~\cite{Juric73} and
the r.m.s.~$d$--$\Lambda$ distance 
$\langle r^2_{d\mbox{-}\Lambda} \rangle^{1/2}=11.2$~fm 
for ${^3_\Lambda{\rm H}}$\,($J^P=1/2^+$, g.s.),
with $\varphi^{(N)}_0$ from three-body $NNN$ calculations 
and $C^2_{TS}=3/2$.

\begin{figure}[t]
\begin{center}
  \includegraphics[width=1.0\linewidth]{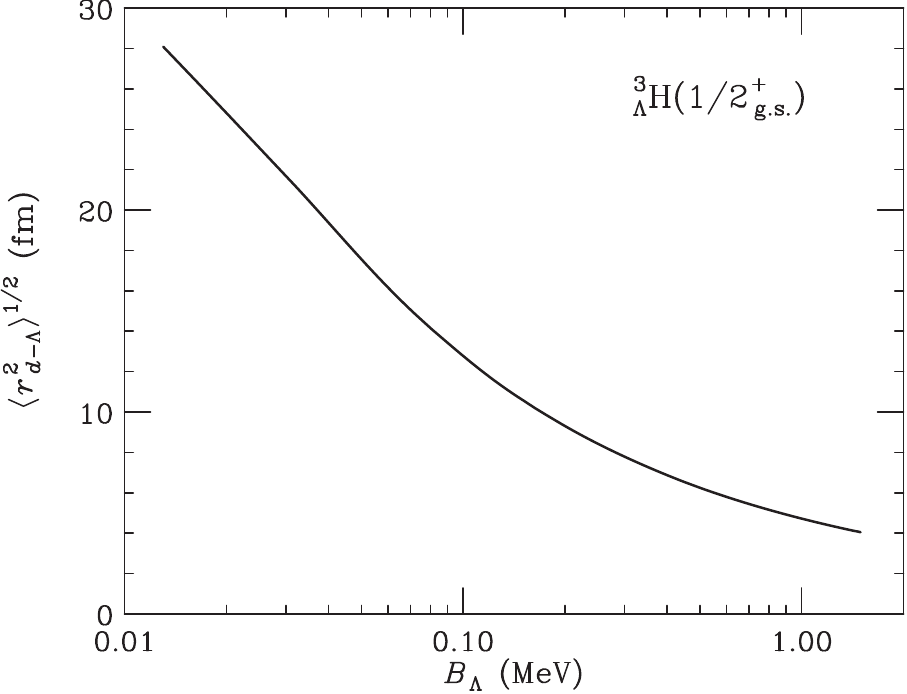}
\end{center}
\caption{\label{fig:rmsBL}
Relation between the $\Lambda$ binding energy $B_\Lambda$ and
the r.m.s.~distance $\langle r^2_{d\mbox{-}\Lambda} \rangle^{1/2}$
for ${^3_\Lambda{\rm H}(1/2^+_{\rm g.s.})}$, illustrating the rapid spatial
expansion of the $d$--$\Lambda$ system toward the weak-binding limit.
}
\end{figure}

The integrated cross section over forward laboratory angles is defined as
\begin{eqnarray}
\sigma_{\rm lab}(^A_\Lambda{\rm Z})\equiv
\int_{\theta_{\rm lab}=0^\circ}^{\theta_{\rm lab}=20^\circ}
\left(\frac{d\sigma}{d\Omega}\right)_{\rm lab,\theta_{\rm lab}}^{J_B^P}d\Omega.
\label{eqn:e6}
\end{eqnarray}
For $^{4}_\Lambda{\rm H}\,(0^+_{\rm g.s.})$, we obtain 
$\sigma_{\rm lab}({^{4}_\Lambda{\rm H}})=$ 63.09~$\mu$b 
at $p_{K^-}=1.0$~GeV/$c$ in DWIA, to be compared with $150.44~\mu$b in PWIA.
Finally, we focus on the ratio of integrated cross sections,
$R_{34}=\sigma_{\rm lab}({^{3}_\Lambda{\rm H}})/\sigma_{\rm lab}({^{4}_\Lambda{\rm H}})$, 
which is designed to retain the genuine $B_\Lambda$ sensitivity
while reducing overall normalization ambiguities common to both systems.

\section{\boldmath
Results and discussion}

\begin{figure}[t]
\begin{center}
  \includegraphics[width=1.0\linewidth]{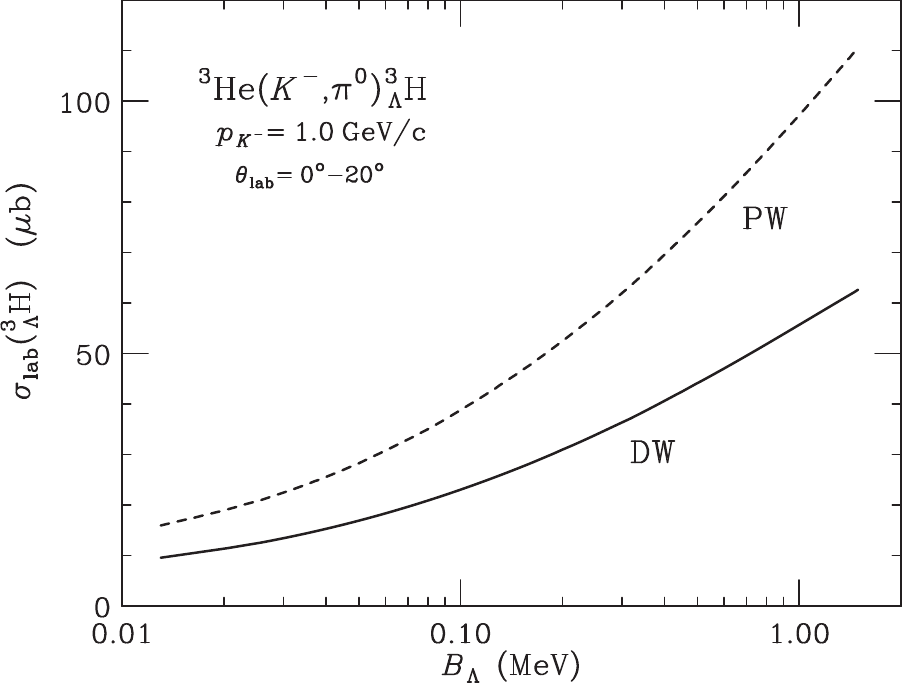}
\end{center}
\caption{\label{fig:csBL}
Calculated integrated cross sections $\sigma_{\rm lab}({^3_\Lambda{\rm H}})$
for ${^3_\Lambda{\rm H}(1/2^+_{\rm g.s.})}$ as a function of $B_\Lambda$.
The solid and dashed curves show the DWIA and PWIA results, respectively.
}
\end{figure}

\begin{figure}[t]
\begin{center}
  \includegraphics[width=1.0\linewidth]{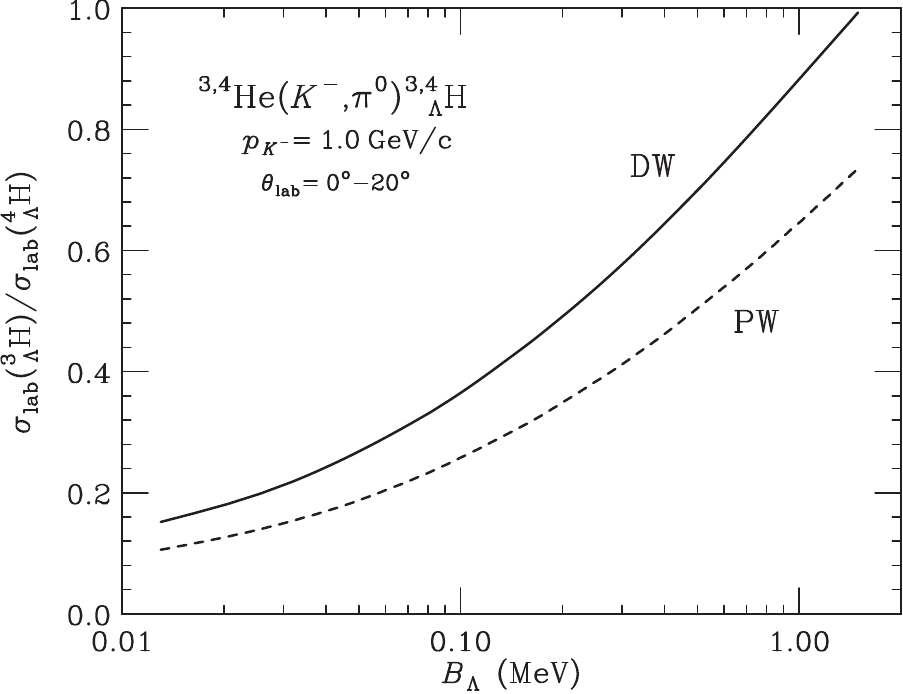}
\end{center}
\caption{\label{fig:R34}
Calculated ratio
$R_{34}=\sigma_{\rm lab}({^{3}_\Lambda{\rm H}})/\sigma_{\rm lab}({^{4}_\Lambda{\rm H}})$
as a function of $B_\Lambda$ for ${^{3}_\Lambda{\rm H}(1/2^+_{\rm g.s.})}$.
The solid and dashed curves show the DWIA and PWIA results, respectively.
}
\end{figure}

\begin{table*}[t]
\caption{
\label{tab:1}
Calculated results of 
the integrated cross sections $\sigma_{\rm lab}({^3_\Lambda{\rm H}})$
for the $^3_\Lambda$H\,($1/2^+_{\rm g.s.}$) bound state in 
the $^{3}$He($K^-$, $\pi^0$) reaction  
at $p_{K^-}=$ 1.0 GeV/$c$
for forward angles $\theta_{\rm lab}=$ 0$^\circ$--20$^\circ$ 
within DWIA and PWIA. 
The ratio 
$R_{34}= \sigma_{\rm lab}
({^3_\Lambda{\rm H}})/\sigma_{\rm lab}({^4_\Lambda{\rm H}})$, 
the $\Lambda$ binding energy $B_\Lambda$, 
the r.m.s.~distance $\langle r^2_{d\mbox{-}\Lambda} \rangle^{1/2}$ 
are obtained, introducing
an artificial strength factor of $f_s$ 
multiplying the $\Lambda$--$d$ interaction.
We use $\sigma_{\rm lab}(^4_\Lambda{\rm H})=$ 63.09~$\mu$b (DWIA) 
and 150.44~$\mu$b (PWIA)~\cite{Harada21}.
}
\begin{ruledtabular}
\begin{tabular}{ccccccc}
\noalign{\smallskip}
$f_s$   & $B_\Lambda$ &   $\langle r^2_{d\mbox{-}\Lambda} \rangle^{1/2}$  
        & \multicolumn{2}{c}{$\sigma_{\rm lab}({^3_\Lambda{\rm H}})$ ($\mu$b)} 
        & \multicolumn{2}{c}{$R_{34}$} \\
\noalign{\smallskip}
\cline{4-5} \cline{6-7}
\noalign{\smallskip}
        &(MeV)& (fm)  & PW  & DW  & PW  & DW \\ 
\noalign{\smallskip}\hline\noalign{\smallskip}
0.900  &0.013  &28.10  & 15.95  & 9.59  &0.106  &0.152 \\
0.925  &0.032  &21.19  & 23.12  &13.84  &0.154  &0.219 \\
0.950  &0.058  &16.43  & 30.31  &18.09  &0.201  &0.287 \\
0.975  &0.091  &13.36  & 37.16  &22.10  &0.247  &0.350 \\
1.000  &0.130  &11.31  & 43.62  &25.85  &0.290  &0.410 \\
1.025  &0.176  & 9.86  & 49.71  &29.37  &0.330  &0.466 \\
1.050  &0.228  & 8.78  & 55.46  &32.66  &0.369  &0.518 \\
1.075  &0.287  & 7.95  & 60.89  &35.75  &0.405  &0.567 \\
1.100  &0.351  & 7.28  & 66.03  &38.64  &0.439  &0.613 \\
1.122  &0.411  & 6.80  & 70.32  &41.05  &0.467  &0.651 \\
1.150  &0.494  & 6.29  & 75.49  &43.92  &0.502  &0.696 \\
1.175  &0.574  & 5.91  & 79.86  &46.32  &0.531  &0.734 \\
1.200  &0.659  & 5.59  & 83.99  &48.59  &0.558  &0.770 \\
1.225  &0.748  & 5.31  & 87.92  &50.72  &0.584  &0.804 \\
1.250  &0.842  & 5.06  & 91.66  &52.73  &0.609  &0.836 \\
1.275  &0.940  & 4.84  & 95.21  &54.62  &0.633  &0.866 \\
1.300  &1.043  & 4.65  & 98.59  &56.41  &0.655  &0.894 \\
1.325  &1.150  & 4.48  &101.80  &58.10  &0.677  &0.921 \\
1.350  &1.260  & 4.32  &104.87  &59.70  &0.697  &0.946 \\
1.375  &1.375  & 4.19  &107.79  &61.21  &0.717  &0.970 \\
1.400  &1.493  & 4.06  &110.58  &62.64  &0.735  &0.993 \\

\noalign{\smallskip}                                                                                                        
\end{tabular}
\end{ruledtabular}
\end{table*}

In evaluating $R_{34}$, we use the production cross section
of $^4_\Lambda$H ($0^+_{\rm g.s.}$) for consistency with the experimental analysis.
Table~\ref{tab:1} summarizes the calculated 
$\sigma_{\rm lab}(^3_\Lambda{\rm H})$ for forward angles 
$\theta_{\rm lab}=$ 0$^\circ$--20$^\circ$ within DWIA and PWIA,
together with the corresponding $B_\Lambda$, 
$\langle r^2_{d\mbox{-}\Lambda} \rangle^{1/2}$,
and $R_{34}$ obtained by 
varying the strength factor $f_s$ of the $\Lambda$--$d$ interaction.
The trend is clear:
as $B_\Lambda$ is reduced, 
the hypernucleus expands significantly and 
the reaction strength reflects this variation.

\subsection{\boldmath 
$B_\Lambda$ dependence of the $^3_\Lambda$H production strength
}

Figure~\ref{fig:rmsBL} shows the relation between $B_\Lambda$ and 
$\langle r^2_{d\mbox{-}\Lambda} \rangle^{1/2}$
in the $^3_\Lambda$H($1/2^+_{\rm g.s.}$) bound state.
For a weakly bound $d+\Lambda$ system, 
the size is not a minor detail but the essential physics:
the long tail of the $d$--$\Lambda$ wave function modifies the overlap in Eq.~(\ref{eqn:e3})
and therefore reshapes the production strength at forward angles.

Figure~\ref{fig:csBL} shows the calculated integrated cross sections 
$\sigma_{\rm lab}(^3_\Lambda{\rm H})$
within the DWIA and PWIA frameworks.
The distortion effect reduces the absolute magnitude of the cross sections,
mainly through absorption of the incoming $K^-$ and outgoing $\pi^0$.
This reduction is not identical for $A=3$ and $A=4$ systems
because of differences in effective nuclear density profiles and recoil kinematics.
Nevertheless, the strong $B_\Lambda$ dependence is not washed out,
since it is governed primarily by the overlap structure in Eq.~(\ref{eqn:e3}).

\subsection{\boldmath 
Cross-section ratio $R_{34}$ and implications for $B_\Lambda$
\label{sec:3B}}

To minimize common normalization uncertainties we consider the ratio
$R_{34}$ defined in Eq.~(\ref{eqn:e0}).
Figure~\ref{fig:R34} shows the calculated values $R_{34}$ 
as a function of $B_\Lambda$.
Because the numerator and denominator are evaluated within the same DWIA framework
and share similar elementary amplitudes and meson distortion,
the dominant normalization effects largely cancel in the ratio.
Consequently, $R_{34}$ provides a more robust observable than the absolute
cross sections while preserving sensitivity to the spatial extension
of the $d$--$\Lambda$ wave function.

We compare our results with the experimental integrated cross sections 
for $\theta_{\rm lab}=0^\circ$--$20^\circ$,
$\sigma^{\rm exp}_{\rm lab}(^3_\Lambda{\rm H}) 
= 15.0 \pm 2.6 \,(\mathrm{stat.})\,^{+2.4}_{-2.8}\,(\mathrm{syst.})~\mu\mathrm{b}$
and
$\sigma^{\rm exp}_{\rm lab}(^4_\Lambda{\rm H}) 
= 49.9 \pm 2.1 \,(\mathrm{stat.})\,^{+7.8}_{-8.0}\,(\mathrm{syst.})~\mu\mathrm{b}$
from the J-PARC E73 experiment~\cite{Akaishi26}. These values yield
\begin{equation}
R_{34}^{\rm exp} = 0.300 \pm 0.054 \,(\mathrm{stat.})\,^{+0.047}_{-0.051}\,(\mathrm{syst.}).
\label{eqn:e7}
\end{equation}
This experimental ratio can be directly compared with the calculated
$R_{34}$ in Fig.~\ref{fig:R34}.
The experimental value $R_{34}^{\rm exp}\approx 0.30$ 
intersects the DWIA curve in the region $B_\Lambda \approx 0.05$--$0.15$~MeV. 
The difference between the DWIA and PWIA results mainly reflects
the effect of meson distortion and recoil kinematics.
We therefore regard it only as a rough indicator of the theoretical
uncertainty associated with the reaction mechanism.
Within this limitation, values of $B_\Lambda$ significantly larger
than 0.3~MeV appear to be disfavored in the present framework.

Using the DWIA curve as the baseline and the PWIA curve as an indicator of model dependence,
the data favor a loosely bound $\Lambda$ hypernucleus with an extended $\Lambda$ distribution.
A sharper constraint will require improved few-body wave functions
and dedicated measurements of the ratio under closely matched kinematics.
Nevertheless, the present result already demonstrates that
the $^3$He($K^-$,$\pi^0$) reaction provides a useful probe of $B_\Lambda$.
In this sense, the present result is broadly consistent with
weak-binding values of $B_\Lambda$ extracted from reaction analyses
of the same system.
The STAR Collaboration~\cite{STAR22} and recent pion spectroscopy at MAMI~\cite{Kino25},
however, have reported substantially larger values.
Within the present DWIA framework, such a larger $B_\Lambda$
would imply a more compact $\Lambda$ wave function
and hence a larger $R_{34}$ than the value of $R^{\rm exp}_{34}$ in Eq.~(\ref{eqn:e7}).
A unified understanding of these different determinations
is beyond the scope of this paper.

\subsection{\boldmath 
Effect of the $^4_\Lambda$H binding energy}

The extraction of the $\Lambda$ binding energy of $^3_\Lambda$H
in the present analysis relies on the theoretical calculation of
$R_{34} = 
\sigma_{\rm lab}(^3_\Lambda\mathrm{H})/\sigma_{\rm lab}(^4_\Lambda\mathrm{H})$,
as discussed in Sec.~\ref{sec:3B}.
In those calculations, the $\Lambda$ binding energy of
$^4_\Lambda$H was taken to be 2.16 MeV,
following the A1 Collaboration result~\cite{Schulz16}.

More recently, the STAR Collaboration~\cite{STAR22} reported 
$B_\Lambda(^4_\Lambda\mathrm{H})=
2.22 \pm 0.06\,({\rm stat.}) \pm 0.14\,({\rm syst.})$~MeV,
while the J-PARC E07 Collaboration~\cite{Kasagi25} reported
$B_\Lambda(^4_\Lambda\mathrm{H})=
2.25 \pm 0.10\,({\rm stat.}) \pm 0.06\,({\rm syst.})$~MeV.
To estimate the sensitivity of the present analysis
to this input parameter,
we have repeated the DWIA calculation of the
$^4_\Lambda$H production cross section
in the forward angles $\theta_{\rm lab}=$ $0^\circ$--$20^\circ$
using $B_\Lambda(^4_\Lambda\mathrm{H})=2.16$ MeV and 2.25 MeV, respectively.
The resulting integrated cross sections are
63.1~$\mu$b for $B_\Lambda(^4_\Lambda\mathrm{H})=2.16$ MeV
and 63.9~$\mu$b for $B_\Lambda(^4_\Lambda\mathrm{H})=2.25$ MeV,
with the corresponding 
$\langle r^2_{t\mbox{-}\Lambda} \rangle^{1/2}=$
3.68 fm and 3.64 fm.
As expected, a slightly larger binding energy 
$B_\Lambda(^4_\Lambda\mathrm{H})$ leads to
a more compact wave function and hence to
a marginal increase of the production cross section 
$\sigma_{\rm lab}(^4_\Lambda\mathrm{H})$.
The change amounts to only about 1.3\%.
Consequently, the theoretical value of the ratio
$R_{34}$ is reduced by a factor of 0.987
when the larger binding energy is adopted.
This confirms that the present extraction of
$B_\Lambda(^3_\Lambda\mathrm{H})$
is primarily governed by the spatial extension
of the $d$--$\Lambda$ wave function,
while the uncertainty associated with the input
$^4_\Lambda$H binding energy plays only a minor role.

\section{Summary and conclusion}
\label{sect:6}

We have studied the production of $^3_\Lambda$H and $^4_\Lambda$H in the 
$^{3,4}$He($K^-$,$\pi^0$) reactions at $p_{K^-}=1.0$~GeV/$c$
within DWIA using the optimal Fermi-averaged $K^-p\to\pi^0\Lambda$ amplitude.
We have emphasized the usefulness of the integrated cross-section ratio
$R_{34}=\sigma_{\rm lab}(^3_\Lambda{\rm H})/\sigma_{\rm lab}(^4_\Lambda{\rm H})$,
which is expected to be stable against common normalization uncertainties.

In conclusion, the production strength of $^3_\Lambda$H and 
the ratio $R_{34}$ exhibit 
strong sensitivity to $B_\Lambda$, 
driven by the spatial extension of the $d$--$\Lambda$ wave function.
Comparison of the calculated ratio $R_{34}$ with the experimental data 
from the J-PARC E73 experiment 
indicates that the $\Lambda$ binding energy of $^3_\Lambda$H is constrained to be
approximately 0.05--0.15~MeV in the present DWIA framework,
supporting the picture of an extremely loosely bound $\Lambda$ hypernucleus.

\begin{acknowledgments}
The authors thank Dr.~Y.~Ma and Dr.~F.~Sakuma for valuable discussions. 
One of the authors (T.~H.) is particularly grateful to Dr.~T.~Akaishi 
for stimulating discussions, which motivated the present study.
\end{acknowledgments}


\clearpage

\end{document}